\documentclass[aps,prl,reprint,showpacs,superscriptaddress]{revtex4-1}

\usepackage[english]{babel} 
\addto\captionsenglish{}

\usepackage{graphicx}
\usepackage{dcolumn}
\usepackage{bm}

\usepackage{xcolor}
\usepackage{comment}
\usepackage{siunitx}

\begin{document}

\preprint{APS/123-QED}

\title{Raman Scattering Cross Section of Confined Carbyne}

\author{Cla Duri Tschannen}
\affiliation{Photonics Laboratory, ETH Zürich, 8093 Zürich, Switzerland}

\author{Georgy Gordeev}
\author{Stephanie Reich}
\affiliation{Department of Physics, Freie Universität Berlin, 14195 Berlin, Germany}

\author{Lei Shi}
\affiliation{School of Materials Science and Engineering, State Key Laboratory of Optoelectronic Materials and Technologies, Nanotechnology Research Center, Sun Yat-sen University, Guangzhou 510275, Guangdong, P. R. China}

\author{Thomas Pichler}
\affiliation{Faculty of Physics, Universität Wien, 1090 Wien, Austria}

\author{Martin Frimmer}
\author{Lukas Novotny}
\affiliation{Photonics Laboratory, ETH Zürich, 8093 Zürich, Switzerland}

\author{Sebastian Heeg}
\email{sebastian.heeg@physik.fu-berlin.de}
\affiliation{Department of Physics, Freie Universität Berlin, 14195 Berlin, Germany}

\date{\today}

\begin{abstract}
We experimentally quantify the Raman scattering from individual carbyne chains confined in double-walled carbon nanotubes. We find that the resonant differential Raman cross section of confined carbyne is on the order of $10^{-22}$ cm$^{2}$\,sr$^{-1}$ per atom, making it the strongest Raman scatterer ever reported.
\end{abstract}

\maketitle


\paragraph{\textbf{Introduction}} Carbyne, an infinitely long linear chain of carbon atoms, is the paradigmatic \textit{sp}-hybridized and truly one-dimensional allotropic form of carbon~\cite{Heimann1999-ya, Hirsch2010-vr}. Due to a Peierls distortion, the equilibrium structure of carbyne is polyynic with alternating single and triple bonds, as shown in Fig. 1(a), and therefore characterized by a dimerized (bi-atomic) unit cell. As a consequence, polyynic carbyne is a direct band gap semiconductor and has one Raman active phonon mode, termed C-mode, which corresponds to an in-phase stretching of the triple bonds along the chain.~\cite{Milani2008-ts, Milani2009-oa}.

Attempts to synthesize carbyne have long been limited by its extreme chemical instability~\cite{Chalifoux2010-ea}. This obstacle has been overcome by growing carbon chains inside carbon nanotubes, which act as nanoreactors and shield the chains from interaction with the environment~\cite{Zhao2003-iz, Fantini2006-wu, Jinno2006-qf, Shi2011-sj, Andrade2015-ml, Zhang2018-ek, Toma2019-tb, Shi2017-zk}. Lengths up to several thousand atoms have been reported for linear carbon chains synthesized inside double-walled carbon nanotubes (DWCNTs)~\cite{Shi2016-oj}. In contrast to short chains comprising few or few tens of atoms~\cite{Wanko2016-mp,Shi2017-zk,Casari2016-so,Casari2018-mz}, the properties of these long chains do not exhibit any length-dependence, indicating that they are the finite realization of carbyne~\cite{Heeg2018-ACS}.

As for other carbon nanostructures~\cite{Thomsen2007-hz, Jorio2011-zt, Ferrari2013-wb, Jorio2017-oa}, Raman spectroscopy is a powerful tool to study the vibrational and electronic properties of carbyne. For carbyne chains encapsulated inside DWCNTs, the C-mode Raman shift and the carbyne band gap energy $E_{\mathrm{gap}}$ are both linearly related to the diameter of the encasing inner nanotube and therefore tunable by choice of nanotube host~\cite{Heeg2018-ACS}. Moreover, strong resonant enhancement of the C-mode Raman signal occurs for excitation energies $E_{\mathrm{L}}$ in the neighborhood of $E_{\mathrm{gap}}$~\cite{Shi2017-zk, Heeg2018-Carbon}.

While Raman scattering is in general a weak effect, strong Raman intensities are predicted by effective conjugation coordinate theory for collective oscillations along the molecular backbone of $\pi$-conjugated systems characterized by a bond-alternation pattern~\cite{Castiglioni1988-gr, Gussoni1991-mb, Castiglioni2004-yt}. For carbyne, this is supported by the fact that even short chains comprising a mere few hundred atoms can be readily detected with conventional Raman spectroscopy. Quantifying the Raman cross section of carbyne is therefore of significant interest due to its anticipated large magnitude, and will enable Raman spectroscopy to be used as a simple yet effective method to evaluate the currently unknown yield of carbyne synthesis directly from far-field bulk measurements~\cite{Shi2016-oj}. Besides, absolute cross sections for Raman scattering can serve as a stringent test to the theories used to describe the scattering mechanism, and it is customary to deduce from these absolute values quantitative results about other core material properties such as electron-phonon coupling constants~\cite{Cardona1979, Grimsditch1981-lr}. Yet, to date, the absolute magnitude of the Raman response of carbyne has remained unexplored.

In this Letter, we experimentally quantify the Raman scattering from confined carbyne. We find that the resonant differential Raman cross section of confined carbyne is on the order of $10^{-22}$ cm${^2}$sr$^{-1}$ per atom, exceeding that of any other known material by two orders of magnitude or more. Our results therefore establish confined carbyne as the strongest Raman scatterer ever reported.

\begin{figure}[t]
\includegraphics[width=\linewidth]{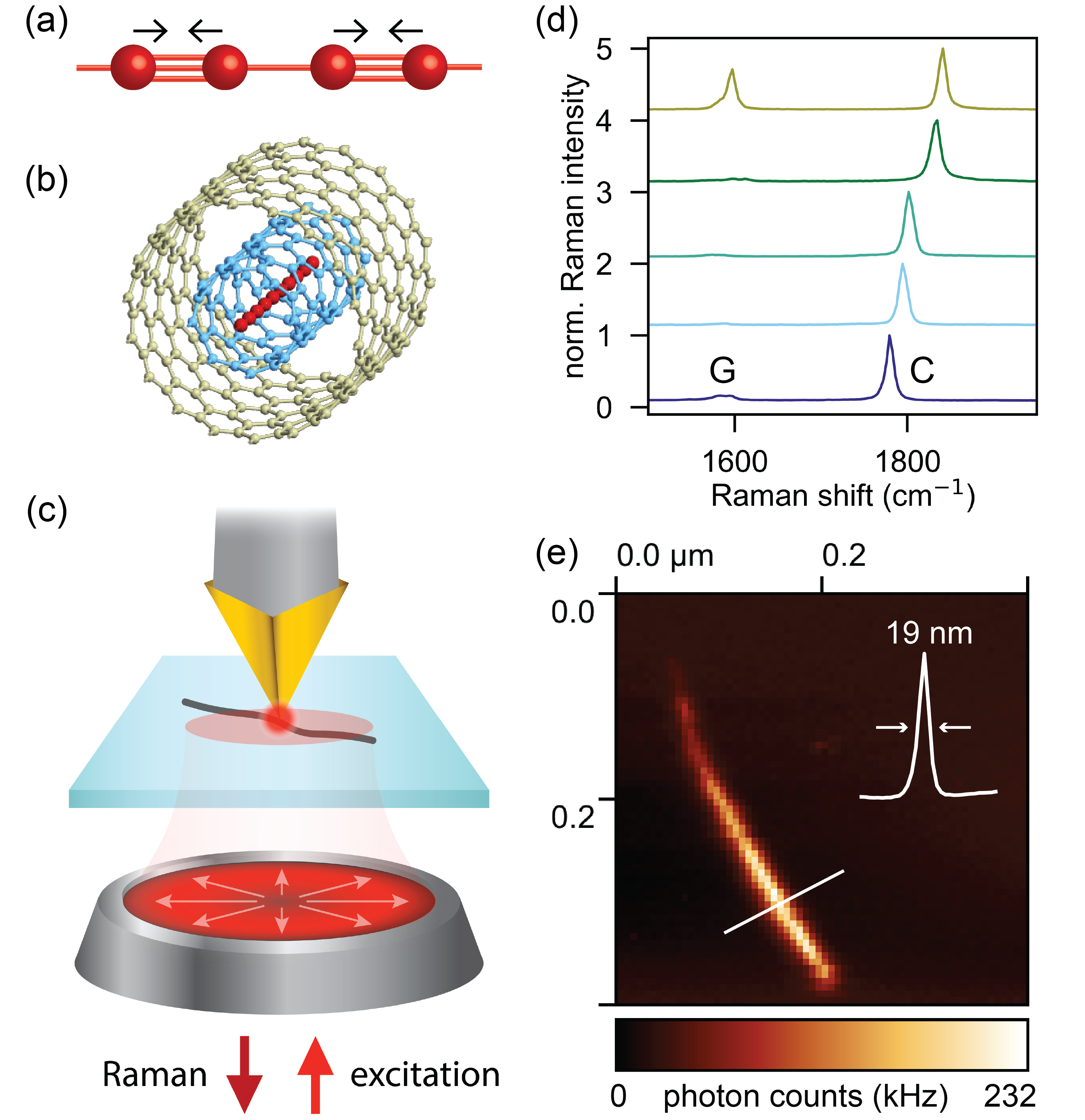}
\caption{\label{fig:Fig1}(a) Atomic structure of polyynic carbyne. Arrows indicate atomic displacements for the Raman-active C-mode. (b) Carbyne chain encapsulated inside a double-walled carbon nanotube (DWCNT). (c) Schematic illustration of our TERS setup. (d) TERS spectra of confined carbyne. The G-peaks arise from the DWCNT. The spectra are vertically offset for better visibility and normalized to the height of the respective C-peak. (e) TERS image of the C-mode of a confined carbyne chain. The intensity profile (inset) extracted along the white line shows a spatial resolution of 19 nm.}
\end{figure}

\paragraph{\textbf{Experimental}} Carbyne chains are grown inside DWCNTs in a high-temperature, high-vacuum process as described in Ref.~\cite{Shi2016-oj} and dispersed on a thin glass coverslip. Our strategy to quantify the resonant differential Raman cross section of confined carbyne then comprises the following steps. First, we identify individual chains by tip-enhanced Raman spectroscopy (TERS) and extract their length for later normalization. Second, we perform far-field Raman measurements with a tunable excitation source to excite each carbyne chain on resonance. Finally, we obtain the absolute differential Raman cross section of carbyne by comparison to a reference scatterer. 

In the first characterization step, we perform TERS using a home-built setup, schematically illustrated in Fig.~\ref{fig:Fig1}(c). A tightly focused radially polarized laser beam (excitation energy 1.96 eV) generates a nanoscale excitation source at the apex of an optical nanoantenna~\cite{Vasconcelos2018-kt}. The antenna is attached to a piezo-electric tuning fork and positioned in close proximity to the sample surface with a shear-force feedback system. The backscattered light is detected either by an avalanche photodiode (APD) or a CCD-equipped spectrometer. In the first case (detection by the APD), the scattered beam passes through a narrow band-pass filter that transmits only the spectral region of carbyne's C-mode. Images are formed by raster-scanning.

\begin{figure}[b]
\includegraphics[width=\linewidth]{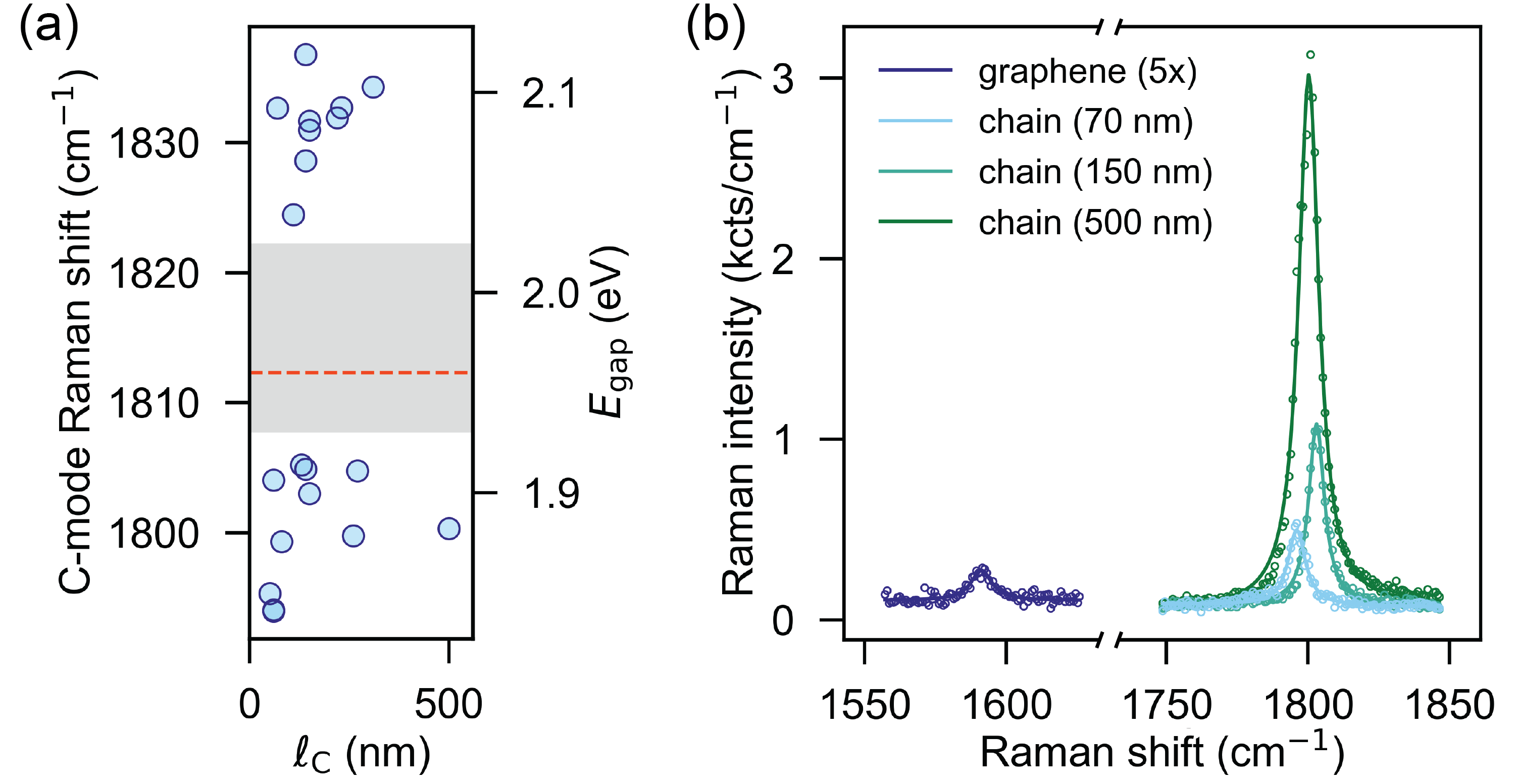}
\caption{\label{fig:Fig2} (a) C-mode Raman shift, band gap energy $E_{\mathrm{gap}}$, and length $\ell_{\mathrm{C}}$ of 20 confined carbyne chains characterized by TERS. The grey area highlights the apparent lack of polyynic carbyne inside metallic host nanotubes (see main text); the dashed line marks the excitation energy (1.96 eV) used for TERS. (b) Far-field Raman spectra of graphene and three carbyne chains of different length, all recorded at 1.89 eV with $\sim$360 $\mu$W power and 60 s integration time. The lines represent Lorentzian fits on a linear background.}
\end{figure}

We consider in this work a total of 20 different confined carbyne chains. For each individual chain, we record a TERS spectrum and a TERS image, from which we obtain the C-mode frequency and the chain length, respectively. Exemplary measurements, recorded with a focal power of $\sim$100 $\mu$W, are displayed in Figs.~\ref{fig:Fig1}(d,e). An overview of all the extracted chain lenghts $\ell_{\mathrm{C}}$ and C-mode frequencies is given in Fig.~\ref{fig:Fig2}(a). No dependence of the C-mode frequency on chain length is evident, indicating that the confined carbon chains are indeed the finite realization of carbyne. In addition, we indicate in Fig.~\ref{fig:Fig2}(a) the band gap energy $E_{\mathrm{gap}}$ for every carbyne chain, which can be directly inferred from the corresponding C-mode frequency~\cite{Shi2017-zk}. The lack of C-mode frequencies between 1808 and 1823 cm$^{-1}$, highlighted by the grey area, is in agreement with other works~\cite{Zhao2003-iz, Fantini2006-wu, Jinno2006-qf, Shi2011-sj, Andrade2015-ml, Zhang2018-ek, Toma2019-tb, Shi2017-zk, Shi2016-oj, Heeg2018-ACS} and explained in Ref.~\cite{Heeg2018-ACS} by an apparent lack of polyynic carbyne inside metallic host nanotubes.

In the second step of our characterization approach, we carry out resonant far-field Raman measurements on the previously identified carbyne chains. We employ a backscattering geometry with polarized detection parallel to the polarization of the excitation laser. A dye laser serves as tunable excitation source. For every carbyne chain, the excitation energy $E_{\mathrm{L}}$ is chosen to closely match the band gap $E_{\mathrm{gap}}$ to ensure maximum resonance enhancement of the Raman intensity ($|E_{\mathrm{L}} - E_{\mathrm{gap}}|$ below 20 meV and thus small compared to the linewidth of the resonance~\cite{Heeg2018-Carbon}). The linear polarization of the excitation beam is adjusted for every measurement to coincide with the chain axis. This takes into account that for one-dimensional systems, owing to their anisotropic polarizability, Raman scattering arises only from the projection of the incoming light onto the system's main axis~\cite{Thomsen2007-hz,Heeg2018-Carbon}.

Figure~\ref{fig:Fig2}(b) shows the Raman spectra of three different carbyne chains excited close to resonance with excitation energy 1.89 eV. We fit each spectrum with a Lorentzian on a linear background. The integrated C-peak intensities obtained from the fits represent the corresponding resonant Raman signal $A_{\mathrm{C}}$ for every chain. We observe that the Raman signal increases with chain length. To extend this finding to carbyne chains with different band gaps, measurements carried out at different excitation energies need to be compared. For this, it is essential to take into account the pronounced wavelength-dependence of the detection system response. 

In our final characterization step we compare the resonant Raman signals of carbyne chains with different band gap energies and calibrate the signal strength. To this end, we employ the method of sample substitution~\cite{Grimsditch1981-lr, Cardona1979}. This consists in normalizing the Raman signal of each individual carbyne chain to that of a substitutional scatterer, for which the absolute differential Raman cross section as a function of excitation energy is known. Here, we choose as a reference the integrated G-peak intensity of graphene, $A_{\mathrm{G}}$ (see Fig.~\ref{fig:Fig2}(b)). We mechanically exfoliate graphene directly onto a glass substrate and identify a monolayer region that is much larger than the laser spot size using hyperspectral Raman mapping (not shown). After every \mbox{(near-)}resonant measurement of a carbyne chain, we acquire the Raman spectrum of graphene using the exact same experimental configuration. In this way, the ratio $A_{\mathrm{C}}/A_{\mathrm{G}}$ of the corresponding signals is deconvolved from the wavelength-dependent response of the detection system. Note that for any given excitation energy we neglect the difference in sensitivity of our detection system for the C-mode of carbyne and the G-mode of graphene. This is only justified because the two modes have very similar Raman frequencies. The laser power before the back aperture of the objective is kept below 400 $\mu$W to avoid sample heating.

\begin{table*}[t]
\caption{Differential Raman cross section $\beta$ of confined carbyne compared to other materials and molecules, determined at excitation energy $E_{\mathrm{L}}$. All values of $\beta$ refer to one atom ($^*$molecule) \textit{in vacuo}. Applied local field correction factors (see main text) range from 0.4 for rhodamine 6G to 39 for carbon nanotubes. }
\begin{ruledtabular}\label{tab:table1}
\begin{tabular}{lccrc}

  & Raman mode (cm$^{-1}$) & $E_{\mathrm{L}}$ (eV) & $\beta$ (cm$^{2}$\,sr$^{-1}$)  & Ref. \\ \hline \rule{0pt}{2.5ex}Confined Carbyne & 1793--1837 & 1.84--2.11 & $(0.7\text{--}1.2)\times 10^{-22}$ & this work \\
  Rhodamine 6G & 1504 & 2.33 & $^*9.6 \times 10^{-25}$ &\cite{Shim2008-ao} \\
  Carbon Nanotube & 200 & 1.96 & \footnote{calculated based on the largest value stated in Ref. \cite{Bohn2010-sr} and assuming a radial breathing mode (RBM) frequency of $200 \text{ cm}^{-1}$ (i.e., $\lambda_{\ell} \approx 105$ nm$^{-1}$~\cite{Maultzsch2005-qz})}$1.7 \times 10^{-25}$ & \cite{Bohn2010-sr} \\
 Graphene & 1584 & 2.41 & $1.3 \times 10^{-27}$ & \cite{Narula2010-ty} \\
 Buckminsterfullerene (C$_{60}$) & 1469 & 1.65 & $^*2.1 \times 10^{-29}$ & \cite{Lorentzen1997-dd} \\
 Diamond & 1332 & 2.41 & $3.4 \times 10^{-30}$ & \cite{Grimsditch1975-fe} \\
 N$_2$ gas & 2331 & 2.54 & $^*3.3 \times 10^{-31}$ & \cite{Fenner1973-af}\\
\end{tabular}
\end{ruledtabular}
\end{table*}

\begin{figure}[b]
\includegraphics[width=\linewidth]{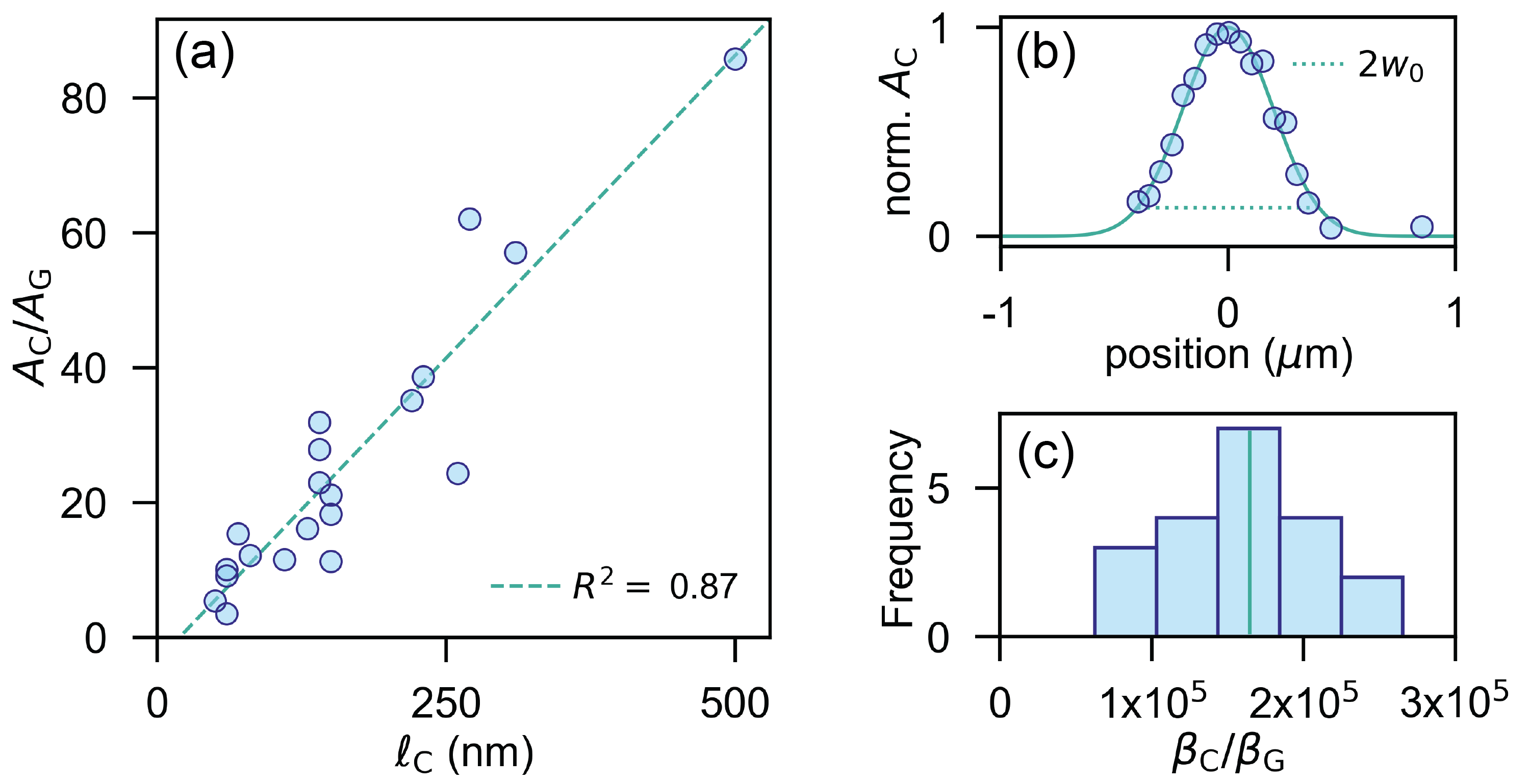}
\caption{\label{fig:Fig3} (a) Resonant Raman signal $A_{\mathrm{C}}$ of confined carbyne, normalized by the G-peak Raman signal of graphene $A_{\mathrm{G}}$, as a function of carbyne chain length $\ell_{\mathrm{C}}$. The linear fit (dashed line) is characterized by an $R^2$ value of 0.87. (b) Profile of the intensity distribution in the focal plane, recorded at 1.84 eV. Gaussian fitting yields a beam waist of $w_0 = 390$ nm. (c) Histogram of the ratio of the differential Raman cross section per atom of carbyne ($\beta_{\mathrm{C}}$) and graphene ($\beta_{\mathrm{G}}$), for 20 different carbyne chains. The green line marks the median value.}
\end{figure}

\paragraph{\textbf{Results and discussion}} The combined results from TERS and resonant far-field Raman measurements on $20$ isolated confined carbyne chains are displayed in Fig.~\ref{fig:Fig3}(a). We find that the resonant Raman signal $A_{\mathrm{C}}$ of carbyne, normalized by the G-peak Raman signal $A_{\mathrm{G}}$ of graphene recorded under the exact same experimental conditions, is proportional to the carbyne chain length $\ell_{\mathrm{C}}$. Moreover, for the longest measured carbyne chains, the Raman signal is almost two orders of magnitude stronger compared to that of graphene. This result is particularly striking when considering that the number of atoms contributing to the respective signals differs substantially. 

Before discussing the magnitude of carbyne's Raman response in more detail, let us first briefly elaborate on the linear length-dependence. As may be observed in Fig.~\ref{fig:Fig3}(a), the linear trend does not start exactly at the origin. This is an artifact of the finite TERS resolution, which leads to a slight overestimation of the carbyne chain lengths $\ell_{\mathrm{C}}$. Residual deviations from the linear fit arise from finite differences $|E_{\mathrm{L}} - E_{\mathrm{gap}}|$ between excitation and carbyne band gap energies and power fluctuations or drift during the signal acquisition. The linear scaling of the carbyne Raman signal with chain length indicates that the phonon coherence length is negligible compared to the chain length and that Raman emission from carbyne can be regarded as an incoherent process~\cite{Cardona1982-le}. Clearly, however, this proportionality is only valid as along as the chain length does not exceed the focal spot size. We verified that this applies to all carbyne chains considered in this work by measuring the waist of the excitation profile in the focal plane. To do so, we scanned a carbyne chain along the direction perpendicular to its main axis through the focal spot and recorded a Raman spectrum every 50 nm. The beam waist $w_0$ can then be extracted from a Gaussian fit to the integrated C-mode intensities $A_{\mathrm{C}}$, as shown in Fig.~\ref{fig:Fig3}(b). Extrapolating $w_0$ to the whole range of excitation energies used for the resonant Raman measurements (1.84--2.11 eV) confirms that even the longest measured carbyne chains displayed in Fig.~\ref{fig:Fig3}(a) lie well within the focal spot. 

We now investigate more closely the magnitude of the Raman response of confined carbyne and how it compares to that of graphene. Since our aim is to quantitatively compare Raman scattering from two materials of different dimensionality, it is instructive to refer to a differential Raman cross section $\beta$ \textit{per atom}, which constitutes an intrinsic property irrespective of spatial extent. We therefore normalize the recorded Raman signals of carbyne and graphene depicted in Fig.~\ref{fig:Fig3}(a) by the corresponding number of atoms $N$ involved in each process. The validity of this approach rests upon the proportionality relation established in Fig.~\ref{fig:Fig3}(a). For carbyne, the number of atoms is simply given by the chain length $\ell_{\mathrm{C}}$ as measured by TERS and the atomic line number density $\lambda_{\mathrm{\ell}} = 7.8$ nm$^{-1}$~\cite{Kastner1995-lp} as $N_{\mathrm{C}} = \ell_{\mathrm{C}}\lambda_{\ell}$. Similarly, for graphene, $N_{\mathrm{G}} = A_{\mathrm{eff}}\rho_{\mathrm{A}}$. Here, $\rho_{\mathrm{A}} = 38.2$ nm$^{-2}$ is the atomic surface number density of graphene~\cite{Baskin1955-cw, Zhao1989-xr}, and $A_{\mathrm{eff}} = \pi w_0^2/2$ denotes the effective surface area of the Gaussian excitation~\cite{Le_Ru2007-ry}. Finally, we can calculate the ratio of the differential Raman cross section per atom of carbyne compared to graphene as $\beta_{\mathrm{C}}/\beta_{\mathrm{G}} = (A_{\mathrm{C}}/N_{\mathrm{C}}) / (A_{\mathrm{G}}/N_{\mathrm{G}})$. 

The results are represented by the histogram in Fig.~\ref{fig:Fig3}(c) for all the measured carbyne chains. We find that the differential Raman cross section per atom of confined carbyne is 5 orders of magnitude larger than that of graphene. Given that graphene is widely recognized as a strong Raman scatterer~\cite{Jorio2011-zt, Ferrari2013-wb}, this result clearly demonstrates a giant Raman effect from confined carbyne. Within our experimental precision, we do not observe a systematic dependence of $\beta_{\mathrm{C}}/\beta_{\mathrm{G}}$ on C-mode frequency. Hence, unlike other properties~\cite{Heeg2018-ACS}, the differential Raman cross section $\beta_{\mathrm{C}}$ of polyynic confined carbyne seems to be largely unaffected by the interaction with the encapsulating nanotube.

Having established a \textit{relative} value for the differential Raman cross section per atom of confined carbyne, we now focus on turning this into an \textit{absolute} result. To this end, we make use of the Raman tensor component $|a| = 92$ Å measured for the G-band of graphene at an excitation energy $E_{\mathrm{L}} = 2.41$ eV in Ref.~\cite{Narula2010-ty}, which translates to a differential Raman cross section per atom of $\beta_{\mathrm{G}} = 1.3 \times 10^{-27}$ cm$^{2}$\,sr$^{-1}$~\cite{Cardona1982-le}. In order to use this value as a reference, we extrapolate it based on the known $E_{\mathrm{L}}^4$-scaling of $\beta_{\mathrm{G}}$~\cite{Cancado2007-gf,Klar2013-im} to the excitation energies (1.84--2.11 eV) used in this work. Together with the median of $\beta_{\mathrm{C}}/\beta_{\mathrm{G}} = 1.6 \times 10^{5}$ shown in Fig.~\ref{fig:Fig3}(c), we find that the differential Raman cross section of confined carbyne amounts to $\beta_{\mathrm{C}} = (0.7\textit{--}1.2) \times 10^{-22}$ cm$^{2}$\,sr$^{-1}$ per atom. The indicated range for $\beta_{\mathrm{C}}$ considers the variation in excitation energy used to resonantly excite carbyne chains with different band gap energies. 

To put our results for the differential Raman cross section per atom of confined carbyne into context, we provide in Table~\ref{tab:table1} a comparison of differential Raman cross sections reported for other carbon allotropes. In addition, we also include Rhodamine 6G, which is known to have a large resonant Raman cross section, and nitrogen gas (N$_2$), which features a weak, non-resonant Raman cross section. Inspection of Table~\ref{tab:table1} underlines that the order of magnitude of the differential Raman cross section of confined carbyne is unprecedented, by far exceeding the reported values for any other material or molecule. We point out that for this comparison to be fair, it is essential to take local field effects into account. The presence of a substrate or solvent modifies the local electromagnetic environment of the sample and thereby affects the measured Raman cross sections~\cite{Boyd1994, Narula2010-ty, Le_Ru2006-at, Le_Ru2007-ry, Le_Ru2008-jq, Etchegoin2010-zv}. We thus applied local field corrections to some of the values listed in Table~\ref{tab:table1}, such that they all represent differential Raman cross sections \textit{in vacuo} and are therefore not obscured by the influence of an underlying or surrounding medium.

\paragraph{\textbf{Conclusion and outlook}} In conclusion, we have demonstrated that confined carbyne is so far the strongest known Raman scatterer, with a resonant differential Raman cross section per atom on the order of $10^{-22}$ cm$^{2}$\,sr$^{-1}$. Its unrivaled Raman cross section renders carbyne a promising candidate for experiments that rely on strong interaction between light and vibrational modes. For instance, combined with plasmonic~\cite{Le_Ru2008-jq} or cavity~\cite{Hummer2016-fs} enhancement, carbyne may provide a platform for studying vibrational strong coupling~\cite{Del_Pino2015-ys}, hot vibrational transitions~\cite{Kneipp1998-pd}, vibrational pumping~\cite{Kneipp1996-ii}, and molecular quantum optomechanics~\cite{Roelli2016-yy}. Further, confined carbyne holds great potential as Raman label for biological imaging. This is similar to dye molecules encapsulated inside functionalized carbon nanotubes, for which such applications have already been demonstrated~\cite{Gaufres2014-ta}. 

However, confined carbyne not only has a differential Raman cross section that is two orders of magnitude larger, but provides the additional advantage that band gap energy and Raman frequency can be tuned by choice of host nanotube~\cite{Heeg2018-ACS, Chimborazo2019-xy}, and that very few other compounds exhibit Raman features in the same spectral region~\cite{McCreery2000}. Finally, we expect that the quantification of carbyne's Raman cross section provided here will serve as a valuable reference point for theoretical calculations, and might start closing the circle into some of the outstanding electronic~\cite{Tongay2004-fk}, thermal~\cite{Wang2015-rl}, and mechanical~\cite{Liu2013-ve} properties anticipated for carbyne. \\

We thank T. L. Vasconcelos of INMETRO for providing the TERS near-field probes and S. Papadopoulos for help with the graphene exfoliation. This work has been supported by the Swiss National
Science Foundation (grant 200020\_192362/1) and by ETH Research (grant ETH-15 19-1). S.H. acknowledges financial support by ETH Z\"urich Career
Seed Grant SEED-16 17-1. L.S. acknowledges the financial support from the National Natural Science Foundation of China (Grant No. 51902353) and Natural Science Foundation of Guangdong Province (Grant No. 2019A1515011227).

\bibliographystyle{apsrev4-1}   
\bibliography{apssamp}

\end{document}